\documentclass[twocolumn]{aastex63}
\usepackage{stfloats}
\usepackage{float}
\usepackage{graphicx}
\usepackage{natbib}
\usepackage{hyperref}
\usepackage{amsmath}
\usepackage{autobreak}
\usepackage{multirow}
\usepackage{makecell}
\usepackage{color}
\usepackage{bm}
\defcitealias{niu2021}{Paper\,{\rm I}}

\received{}
\revised{}
\accepted{}
\submitjournal{ApJL}

\shorttitle{Correction to the photometric colors of \textit{Gaia} EDR3} 
\shortauthors{Niu et al.}

\graphicspath{{./}{figures/}}

\begin{document}

\title{Correction to the photometric colors of \textit{Gaia} Early Data Release 3} 

\correspondingauthor{Haibo Yuan}
\email{yuanhb@bnu.edu.cn}

\author{Zexi Niu}
\affiliation{National Astronomical Observatories,
Chinese Academy of Sciences\\
20A Datun Road, Chaoyang District,
Beijing, China}

\author{Haibo Yuan}
\affiliation{Department of Astronomy,
Beijing Normal University \\
19th Xinjiekouwai Street, Haidian District,
Beijing, China}

\author{Jifeng Liu}
\affiliation{National Astronomical Observatories,
Chinese Academy of Sciences\\
20A Datun Road, Chaoyang District,
Beijing, China}

\begin{abstract}

In this work, we use the spectroscopy-based stellar color regression (SCR) method with $\sim$ 0.7 million common stars between
LAMOST DR7 and \textit{Gaia} EDR3 to acquire color corrections in $G - G_{\rm RP}$ and $G_{\rm BP} - G_{\rm RP}$. 
A sub-mmag precision is achieved. Our results demonstrate that improvements in the calibration process of the EDR3 
have removed the color term in $G_{\rm BP} - G_{\rm RP}$ and eliminated the discontinuity caused by 
the changes of instrument configurations to a great extent. 
However, modest systematic trends with $G$ magnitude are still detected. 
The corresponding color correction terms as a function of $G$ are provided for $9.5 < G < 17.5$ mag and compared with other determinations. We conclude that the corrections given in this work are particularly suited for cases where the color-color investigations are required while for color-magnitude investigations other corrections may be better due to systematic with reddening. Possible applications of our results are discussed.

\end{abstract}


\section{Introduction} \label{sec:intro}

Very recently, \textit{Gaia} Collaboration published the Early Data Release 3 (EDR3) \citep{gaia2020} based on the first 34 months of its nominal mission \citep{gaia2016}, providing $G$ band photometry for 1.8 billion sources brighter than 21 and 1.5 billion sources with $G_{BP}$ and $G_{RP}$ photometry, with an uniform calibration at mmag level. In the \textit{Gaia} DR2 era, several works have detected the magnitude dependent systematic errors up to 10 mmag or higher \citep{maw2018,wei2018,casa2018,niu2021} and the discontinuities caused by the changes of instrument configurations \citep{evans2018,niu2021}. Among them, using about 0.5 million well selected common stars between the LAMOST DR5 \citep{luo2015,zhao2012} and \textit{Gaia} DR2, \citet{niu2021} (hereafter Paper\,{\rm I}) applied the stellar color regression 
(SCR) method \citep{yuan2015} to calibrate the $G - G_{\rm RP}$ and $G_{\rm BP} - G_{\rm RP}$ colors. With an unprecedented precision of about 1 mmag, systematic trends with $G$ magnitude are revealed in great detail for both colors, reflecting changes in instrument configurations. Color-dependent trends are also found for the $G_{\rm BP} - G_{\rm RP}$ for stars brighter than $G \sim$ 11.5 mag. 

From DR2 \citep{gaia2018} to EDR3, a number of important improvements have been implemented to further reduce its photometric (random and systematic) errors, including the fitting of $G$ fluxes, the processing of BP and RP spectra, and the calibration processes \citep{phot_cont,validation}. The median uncertainties of $G$ magnitudes are reduced by almost a factor of two, reaching 0.2 mmag at $10 < G < 14$, 0.8 mmag at $G \sim$ 17, and 2.6 mmag at $G \sim$ 19. 
In order to take full advantage of its exquisite photometric quality, in this work, we follow the same routine of Paper\,{\rm I} to validate and correct possible magnitude/color dependent systematics in EDR3.

The paper is organized as follows. We briefly describe our data and method in Section 2, with differences from Paper\,{\rm I} stressed. The result is presented and discussed in Section 3. We summarize in Section 4.

\section{Data and Method} \label{sec:data}

We use the same method and follow the same steps as in Paper\,{\rm I}, which are briefly summarized below.
Details of criteria used to select samples and the SCR method can be found in Paper\,{\rm I}.

We combine the newest \textit{Gaia} EDR3 with the LAMOST DR7 \citep{luo2015} and apply the same constraints as in Paper\,{\rm I},
e.g., E(B$-$V) $<$ 0.05 mag according to the \citeauthor*{sfd} (\citeyear{sfd}, hereafter SFD) dust reddening map, galactic latitude $|b| > 20$ deg, and vertical distance to the galactic disk $|Z| > 0.2$ kpc. In Figure 3 of \citetalias{niu2021}, we clarified that the signal-to-noise ratio for the $g$ band ($S/N_{\rm g}$) of 20 is sufficient for this work because there are no systematic effects in the LAMOST stellar parameters with the $S/N_{\rm g}$ till at very low $S/N_{\rm g}$ of about 15. So we adopt a lower cut on $S/N_{\rm g}$ of 20 to generate a larger sample. At $S/N_{\rm g}$ $>$ 20, the agreement of the LAMOST parameters with APOGEE is better than 120 K for $T_{\rm eff}$, 0.15 dex for $Log$ g, and 0.1 dex for [Fe/H], as shown in Figure 3 of \citetalias{niu2021}.

We finally select a sample containing 779,691 main-sequence (MS) stars and 71,952 Red Giant Branch (RGB) stars, covering $9.5 < G < 17.5$ magnitude range. 
Note that as proposed by the \citet{gaia2020} and \citet{phot_cont}, the corrected $G$ magnitudes for sources with 6-parameter astrometric solutions are used in this paper.

Due to the different passbands between DR2 and EDR3, their reddening coefficients are slightly different. Following 
Sun et al. (to be submitted), we have empirically determined the temperature and reddening dependent reddening 
coefficients for the EDR3 colors, as given by the Equations \ref{eq1} and \ref{eq2}. Note that the reddening corrections in this work do not take into account the distance of sources and assume that all sources to be beyond than the source of reddening. All colors referred to hereafter are dereddened using the SFD map and the empirical coefficients.

\begin{equation}\label{eq1}
 \begin{aligned}
  &  R(G_{\rm BP}-G) = 1.428 - 0.539 \times E(B-V)_{\rm SFD} + 0.406 \\
  &  \times {E(B-V)}^2_{\rm SFD} - 1.976 \times 10^{-4} \times T_{\rm eff} + 2.004 \times 10^{-5} \\
  &  \times T_{\rm eff}  \times E(B-V)_{\rm SFD} - 1.187 \times 10^{-8} \times {T_{\rm eff}}^2
 \end{aligned} 
\end{equation}

\begin{equation}\label{eq2}
 \begin{aligned}
  &  R(G_{\rm BP}-G_{\rm RP}) = 1.684  - 0.839 \times E(B-V)_{\rm SFD} + 0.555 \\
  &  \times {E(B-V)}^2_{\rm SFD} - 1.390 \times 10^{-4} \times T_{\rm eff} - 3.142 \times 10^{-6} \\
  &  \times T_{\rm eff} \times E(B-V)_{\rm SFD} + 1.520 \times 10^{-8} \times {T_{\rm eff}}^2
 \end{aligned}
\end{equation}

\begin{figure*}[htbp] 
\flushleft
    \includegraphics[width=7.5in]{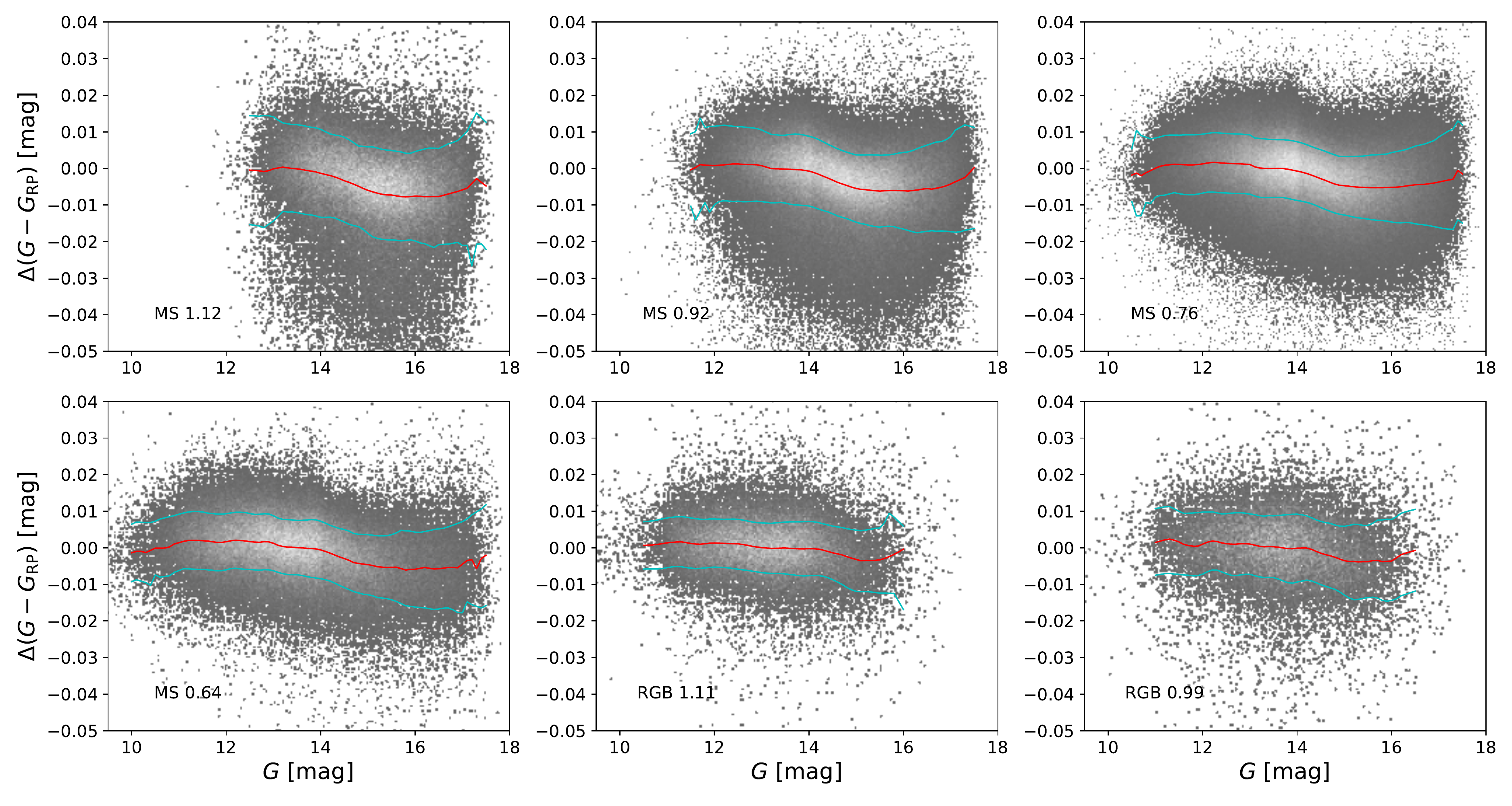} 
    \caption{$G-G_{\rm RP} $ residuals against $G$ magnitude for six subsamples labelled by their evolution stages and median ($G_{\rm BP}-G_{\rm RP}$)$_0$ colors. Red and cyan lines are the LOWESS (locally weighted scatterplot smoothing) of the median values and the standard deviations for the equally spaced $G$ bins, respectively.\label{fig1}} 
\end{figure*}

\begin{figure*}[htbp] 
\flushleft
    \includegraphics[width=7.5in]{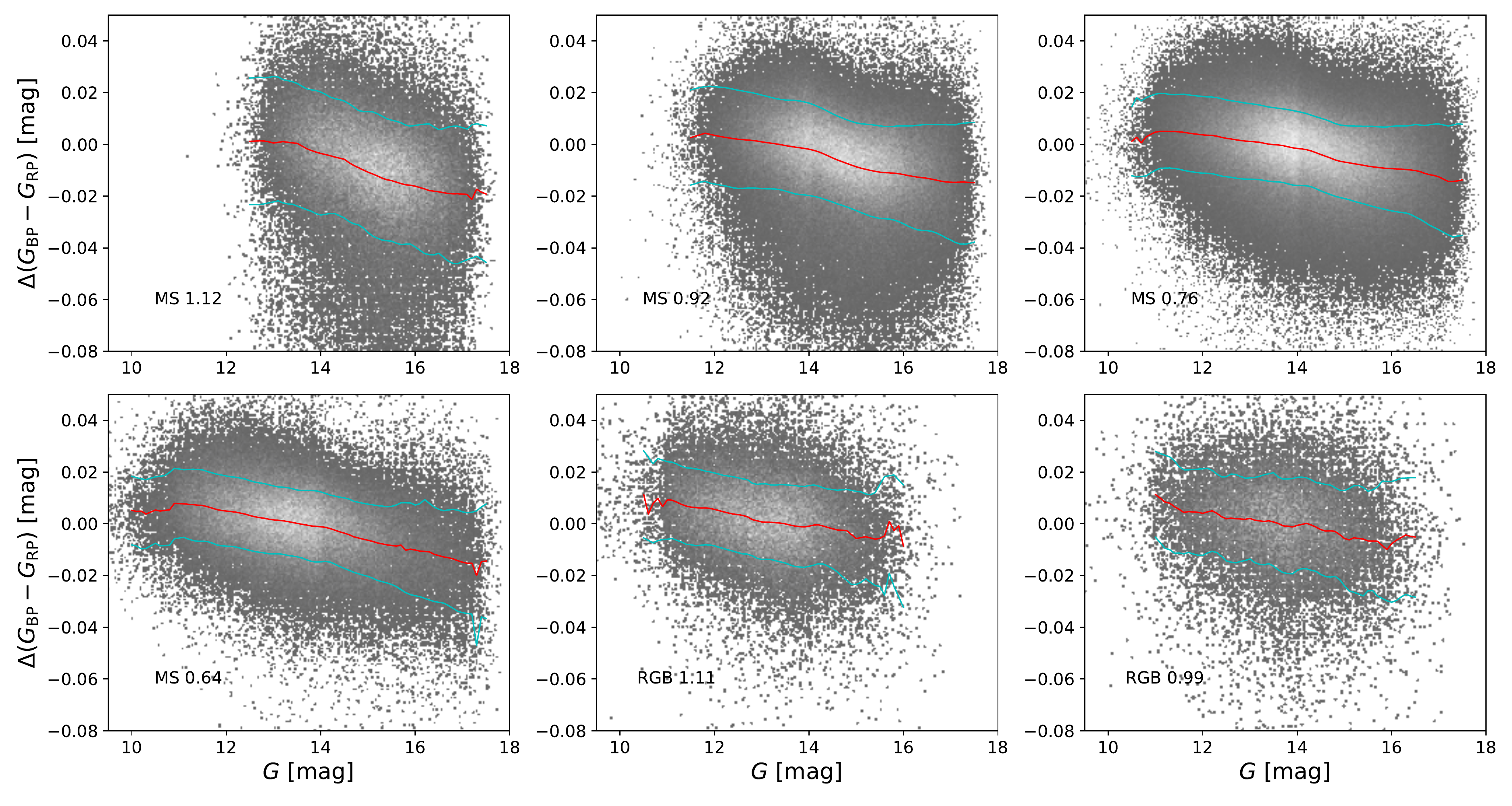} 
    \caption{Same to Figure \ref{fig1} but for $G_{\rm BP}-G_{\rm RP}$ residuals.\label{fig2}} 
\end{figure*}

As in Paper\,{\rm I}, a control sample of $13.3 < G < 13.7$ are selected to define the empirical relations between the \textit{Gaia} intrinsic colors and the LAMOST DR7 stellar parameters. Then we divide the selected samples into six subsamples of different evolution stages and colors. MS stars are divided into four subsamples with the median $G_{\rm BP}-G_{\rm RP}$ colors of 1.12, 0.92, 0.76, and 0.64 mag, respectively. RGB stars are divided into two subsamples, with the median $G_{\rm BP}-G_{\rm RP}$ colors of 1.11 and 0.99 mag, respectively. Applying these relations to a given (sub-)sample, the median values of the color residuals between the SCR-derived colors and the corresponding \textit{Gaia} EDR3 ones as a function of $G$ are obtained as the correction terms, as shown in Figures \ref{fig1} and \ref{fig2}. A 3$\sigma$-clipping is performed in the process. Specifically, corrections can be expressed as:
\begin{equation}\label{eq3}
    C^{'} = C + \Delta C
\end{equation}
where $C^{'}$ is the corrected color, $C$ is the \textit{Gaia} EDR3 color and $\Delta C$ is the color correction term. 

\section{Result and Discussion} \label{sec:methodresult}

\begin{figure}[htbp] 
    \centering 
    \includegraphics[width=3.5in]{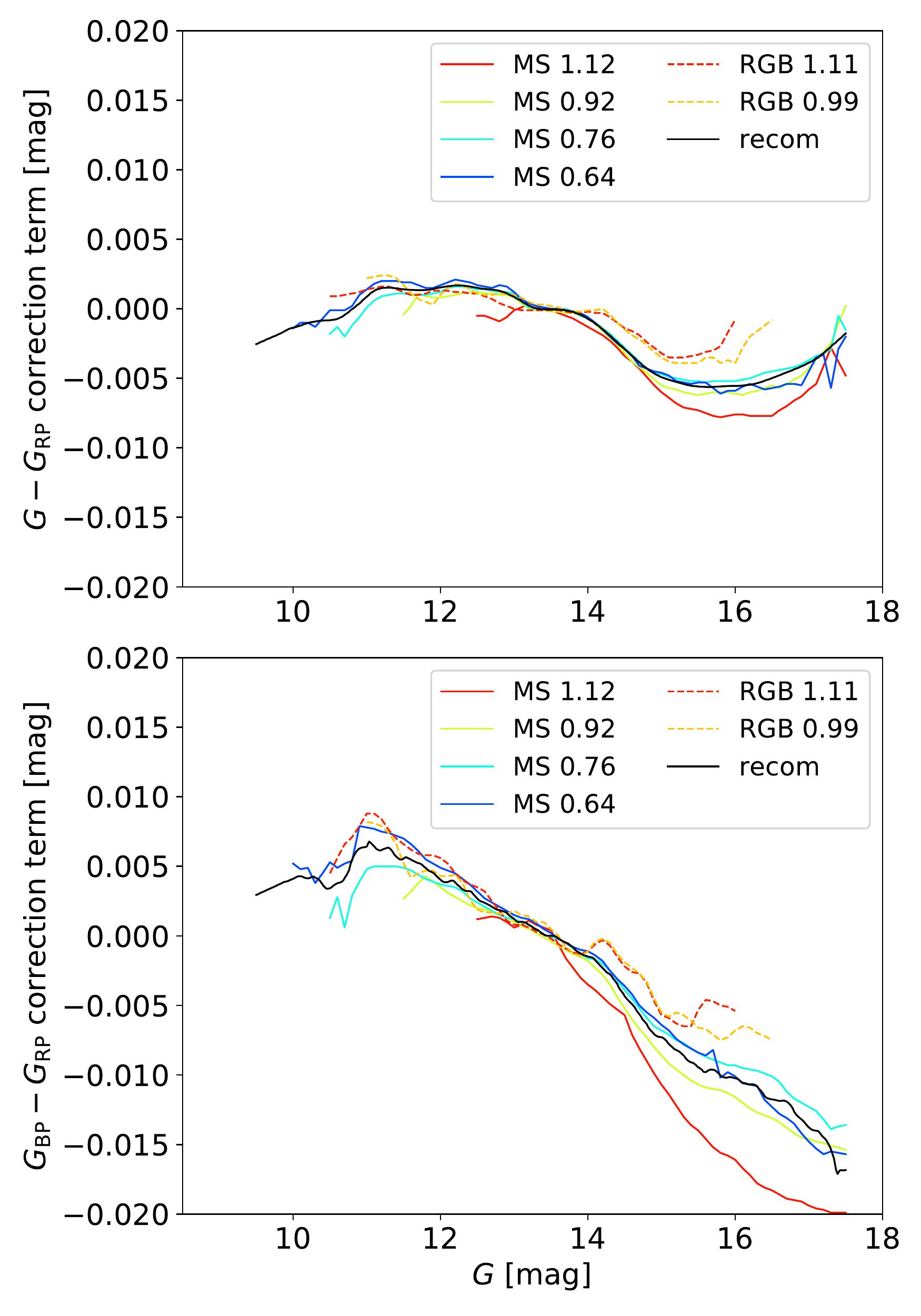} 
    \caption{Color correction curves yielded by different subsamples and the recommended ones for $G-G_{\rm RP}$ (\textit{top}) and $G_{\rm BP}-G_{\rm RP}$ (\textit{bottom}). \label{fig3}} 
\end{figure}

Figure \ref{fig3} plots color correction curves yielded by different subsamples. As already demonstrated in Paper\,{\rm I}: (1) the correction terms are independent of the stellar evolution stages; (2) the inconsistency between the red and blue subsamples when $G > 14$ mag is due to the selection function of the LAMOST data and the spatially dependent systematics of the SFD reddening map. Therefore, we compute the recommended correction curves using the three blue MS subsamples (MS 0.92, MS 0.76, and MS 0.64). The results are over-plotted in black lines in Figure \ref{fig3} and listed in Table \ref{tab:1}. The standard deviations of the difference between the recommended curve and the three blue MS subsamples for $G-G_{\rm RP}$ and $G_{\rm BP}-G_{\rm RP}$ are 0.3 and 0.7 mmag, respectively. This suggests that the calibration curve of each subsample has a typical random error smaller than 1.0 mmag, and the random errors of the recommended curves are even smaller. In the top panel of Figure \ref{fig3}, the systematic trend of $G-G_{\rm RP}$ is smaller than the one found for \textit{Gaia} DR2, with a maximum range of less than 10 mmag in total. Feature coming from the observation modes changing at $G \sim 16$ mag is gone and those at $G \sim$ 11 and 13 mag are visible but at a level of about 1 mmag from the recommended curve. This confirms the big improvement in the $G$ band photometry. The bottom panel shows that the $G_{\rm BP}-G_{\rm RP}$ correction curves are no longer color-dependent at the bright end, at least for F/G/K stars. However, the magnitude-dependent trend is still significant, as in DR2.

\begin{figure*}[htbp] 
    \centering 
    \includegraphics[width=7in]{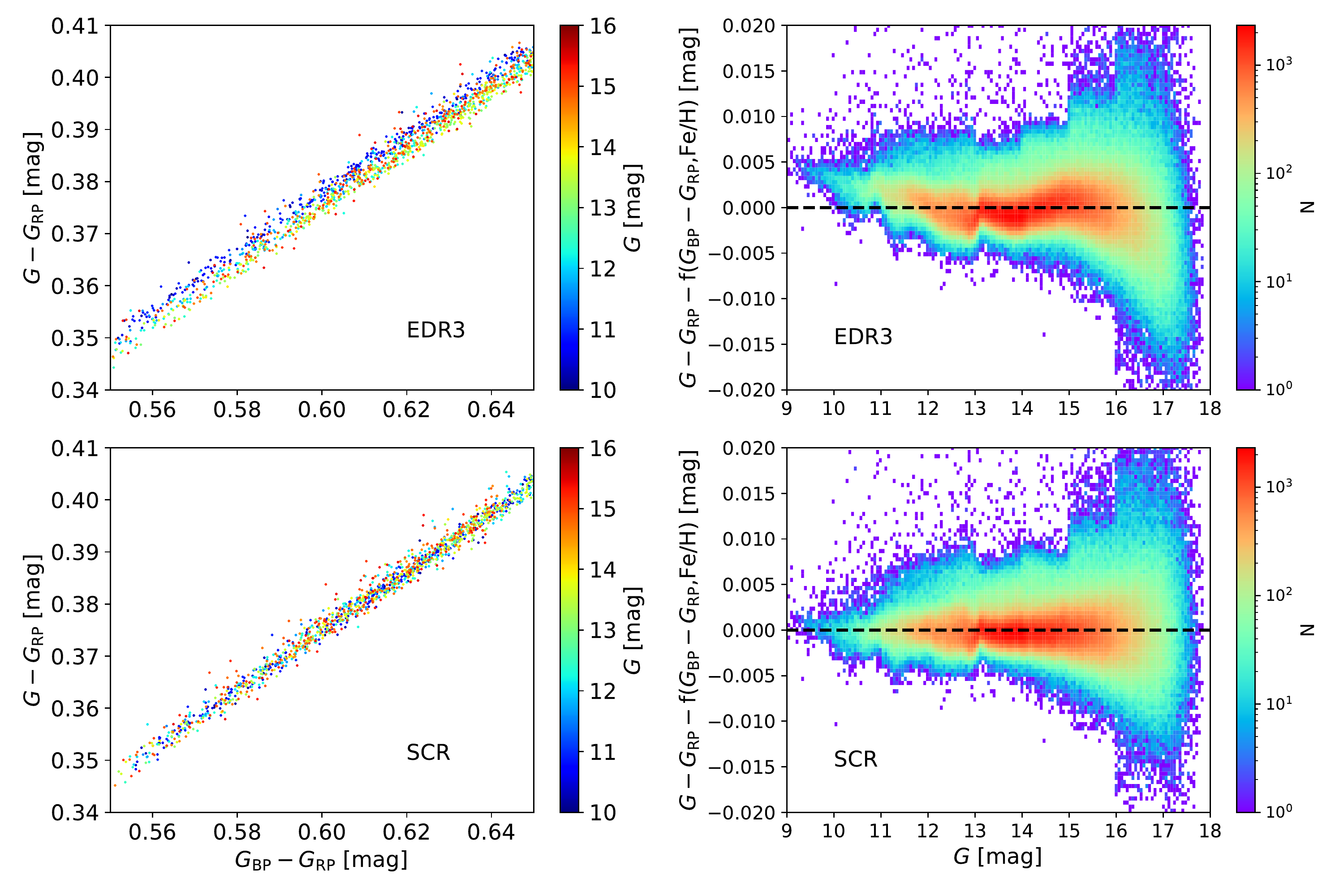}  
    \caption{From \textit{top} to \textit{bottom}: before and after color corrections. \textit{left}: color-color diagram of a MS sample within a narrow [Fe/H] range. \textit{right}: Residuals from a global $G-G_{\rm RP}=f(G_{\rm BP}-G_{\rm RP},{\rm [Fe/H]})$ relation. \label{fig4}} 
\end{figure*}

To test our color correction curves, we select a MS sample within a narrow [Fe/H] range of $-0.5 < {\rm Fe/H} < -0.25$, and compare their distributions in the color-color diagram before and after corrections using the recommended curves in Table \ref{tab:1}. The results are shown in the left panels of Figure \ref{fig4}.
Magnitude-dependent offsets in the color-color diagram are clearly seen from the top left panel.
After corrections, there are no offsets, and the total width becomes narrower as expected. 
Furthermore, like Figure 32 in \citet{validation} of EDR3 and Figure 31 in \citet{arenou2018} of DR2, we plot a 2D histogram of the $G - G_{\rm RP}$ residuals in the right panels of 
Figure \ref{fig4} with respect to the metallicity-dependent stellar color locus. The improved yet still subsistent trend with magnitude is clearly seen in the top panel, especially at the bright and faint ends. After corrections of this work, the trend becomes flat and centered at zero, demonstrating the power of our corrections in the color-color diagram.

\begin{figure}[htbp] 
    \centering 
    \includegraphics[width=3.5in]{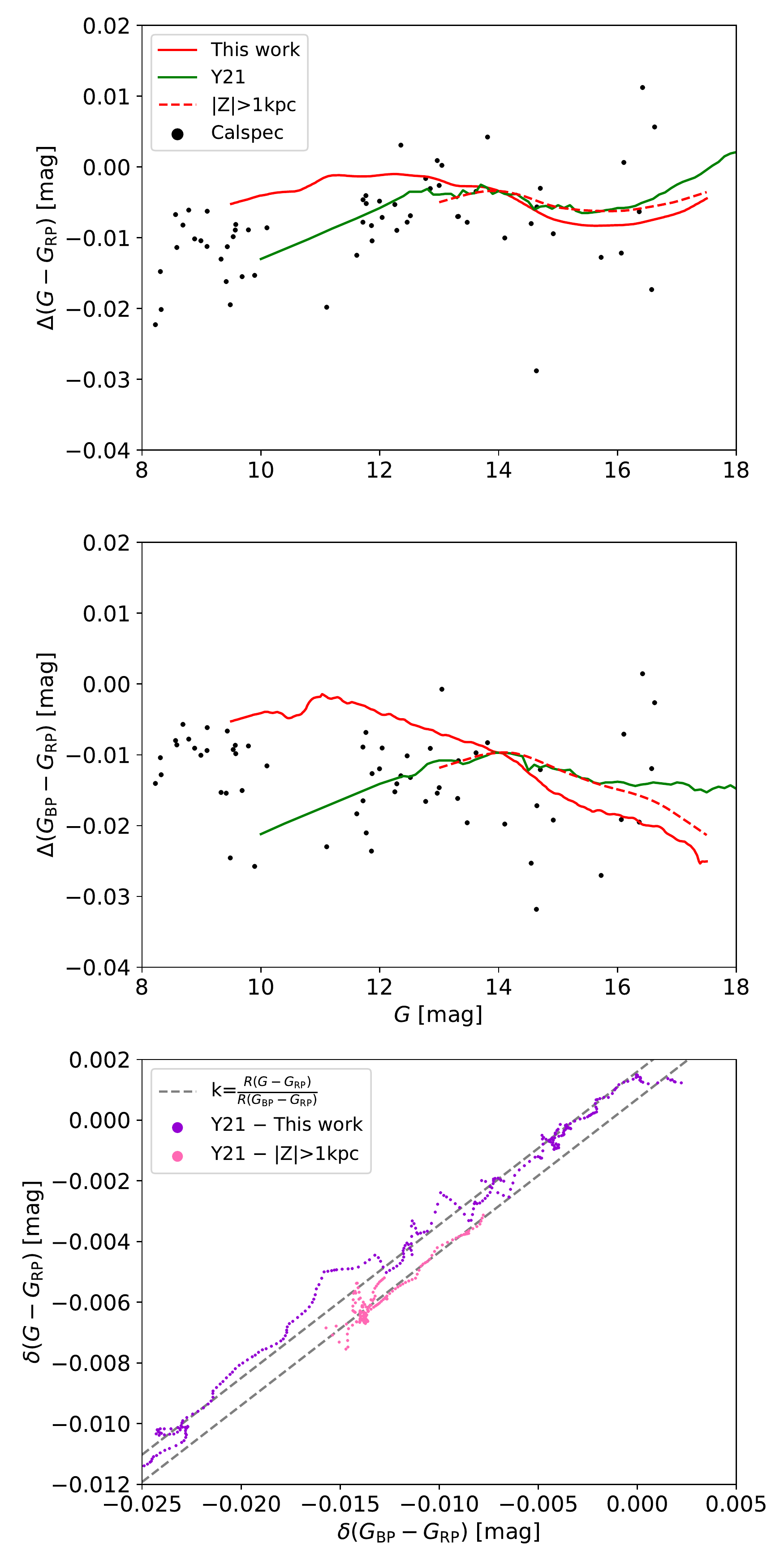}  
    \caption{\textit{Top and middle:} Comparisons of different color correction curves. Red solid line: $|Z| > 0.2$ kpc using the SCR method. Red dashed line: $|Z| > 1$ kpc using the SCR method. Black dots: from the CALSPEC spectra. Green line: from \citet{yl}. Both red solid and dashed lines are systematically shifted by a few mmag to match the green lines at 14 mag for comparison.} \textit{bottom}: Differences of the color correction terms in $G_{\rm BP}-G_{\rm RP}$ versus those in $G-G_{\rm RP}$. The two dashed lines have the same slope, determined by the median value of $\frac{R(G-G_{\rm RP})}{R(G_{\rm BP}-G_{RP})}$.  \label{fig5} 
\end{figure}

There are also color corrections deduced from other approaches. For example, using 10,000 well selected Landolt standard stars \citep{lan2013}, \citet{yl} has obtained magnitude and color corrections of the \textit{Gaia} EDR3 by training the observed $UBVRI$ magnitudes into the \textit{Gaia} EDR3 magnitudes and colors. Their results (red lines in their Figure 4) are plotted in green lines in Figure \ref{fig5}. Results from synthetic colors of the CALSPEC (\citealt{calspec}) spectra are also plotted in black dots. Only 61 stars with \texttt{phot}$\_$\texttt{bp}$\_$\texttt{rp}$\_$\texttt{excess}$\_$\texttt{factor} $< 0.1$ after \citet{phot_cont} correction and $G > 8$ mag are shown. 
Note that \citet{yl} adopted a control sample of $17 < G < 17.5$ mag, their correction curves were shifted to match the results of the CALSPEC spctera. In this work, we have adopted a different control sample of $13.2 < G < 13.6$ mag, therefore, a constant offset may well exist. To make a straightforward comparison, the recommended curves in this work, which are over-plotted in red solid lines, are systematically shifted by a few mmag to match the \citet{yl} corrections at 14 mag.
Given the small number of black dots and their internal scatter (about 10 mmag for $G-G_{\rm RP}$ and 15 mmag for $G_{\rm BP}-G_{\rm RP}$), both results from this work and \citet{yl} are consistent with the black dots. However, discrepancies between this work and \citet{yl} are up to about $\pm$5 mmag for $G-G_{\rm RP}$ and $\pm$10 mmag for $G_{\rm BP}-G_{\rm RP}$, and are correlated with the $G$ magnitude.

To investigate the origins and effects of the discrepancies between this work and \citet{yl}, their differences of the color corrections without shifting in $G_{\rm BP}-G_{\rm RP}$ are plotted against those in $G-G_{\rm RP}$ in the bottom panel of Figure \ref{fig5}. A strong linear correlation is found. The slope is in good agreement with the median value of $\frac{R(G-G_{\rm RP})}{R(G_{\rm BP}-G_{RP})}$. Considering that reddening corrections are not involved in \citet{yl}, the result suggests that the discrepancies are mainly caused by imperfect reddening corrections in this work. We use the 2D SFD reddening map. Despite the systematic errors that depend on spatial position and dust temperature (e.g., \citealt{pg2010}; Sun et al. to be submitted) in the SFD reddening map as discussed in Paper\,{\rm I}, the map also tends to overestimate reddening for stars that are within the Galactic dust layer. Although in this work (and Paper\,{\rm I}) we require stars of $|Z| > 0.2$ kpc, their reddening corrections may be overestimated to some extent around 0.01 mag in E(B-V), depending on their distances/magnitudes. It is not surprising, as there are increasing evidences supporting the co-existence of a thin dust disk and a thick dust disk in the Galaxy (e.g., Yuan et al. in prep.; \citealt{guo}; Zhang et al., to be submitted). The scale height of the thick dust disk is about 200 - 400 pc in the solar neighborhood. Dust clouds in the Galactic halo (Yuan et al. in prep.) and halos of other galaxies (e.g., M\,31 and M\,33, \citealt{zhangdust}) are also detected. We redo the procedure with an excessive sample of $|Z| > 1$ kpc and a control sample of $15.3 < G <15.5$ mag. Its results are plotted in red dashed lines, which match the green lines much better, consistent with the above scenario. It suggests that our correction curves of individual colors are subject to systematic errors from reddening correction using the SFD map. We further test other 2D reddening maps of \citet{lenz2017}, \citet{pg2010}, and \citet{planck}, and find no big differences with the SFD map. 

Fortunately, even though both correction curves of the two \textit{Gaia} colors suffering systematic errors in reddening correction, their systematic errors are largely canceled in the color-color diagram, as the reddening vector is almost parallel to the stellar locus in the color-color diagram. We estimate the standard deviation of the differences between the purple points and their corresponding gray line, which is 0.3 mmag and contributed by the random errors of corrections from \citet{yl} and this work. Given that the typical errors of color correction curves of \cite{yl} are about 0.2 -- 0.4 mmag, the small standard deviation of 0.3 mmag suggests that the typical uncertainties in this work are much smaller than 0.3 mmag. Moreover, with 0.7 million stars, our sample enables a much higher resolution in $G$ magnitude,
yielding improved corrections in the color-color diagram compared to \citet{yl} (See their Figure\,5).

It is worth clarifying that our color corrections in this work and in Paper\,{\rm I} are precise to sub-mmag only in the cases where a color-color diagram is used. 
In other cases, for example where a color-magnitude diagram is used, the corrections from \citet{yl} are preferred. 
Another thing should be mentioned is that we compute the curves within approximately $9.5 < G < 17.5$ and $0.6 < G_{\rm BP}-G_{\rm RP} < 1.3$, so for stars outside the above ranges, e.g., very blue stars brighter than G $\sim$ 13 \citep{phot_cont}, the curves should be used with caution.
Nevertheless, this corrected color-color diagram could be helpful in a number of studies, for example, determining reliable photometric metallicities for an enormous and magnitude-limited sample of stars from \textit{Gaia} (Xu et al. to be submitted) and estimating binary fractions of a volume-limited sample of stars (Niu et al. to be submitted). 

\setlength{\tabcolsep}{1.5mm}{
\begin{table*}[htbp]
\footnotesize
\centering
    \caption{Color correction curves. The first column is $G$ magnitude. The second and third columns are respectively recommended $G-G_{\rm RP}$ and $G_{\rm BP}-G_{\rm RP}$ calibration terms in units of mmag. \label{tab:1}}
    \begin{tabular}{ccc|ccc|ccc|ccc|ccc|ccc} 
     \textbf{G} & \multicolumn{2}{c|}{\textbf{recom}} & 
     \textbf{G} & \multicolumn{2}{c|}{\textbf{recom}} &
     \textbf{G} & \multicolumn{2}{c|}{\textbf{recom}} &
     \textbf{G} & \multicolumn{2}{c|}{\textbf{recom}} &
     \textbf{G} & \multicolumn{2}{c|}{\textbf{recom}} &
     \textbf{G} & \multicolumn{2}{c}{\textbf{recom}} \\
     \hline

 9.50 &  $-$ 2.54 & 2.95 & 10.85 & 0.17 & 6.00 & 12.20 & 1.68 & 3.84 & 13.55 &  $-$ 0.04 &  $-$ 0.02 & 14.90 &  $-$ 4.65 &  $-$ 7.07 & 16.25 &  $-$ 5.41 &  $-$ 10.68 \\ 
 9.53 &  $-$ 2.48 & 3.01 & 10.88 & 0.30 & 6.19 & 12.23 & 1.68 & 3.68 & 13.58 &  $-$ 0.04 &  $-$ 0.10 & 14.93 &  $-$ 4.74 &  $-$ 7.19 & 16.28 &  $-$ 5.38 &  $-$ 10.75 \\ 
 9.56 &  $-$ 2.39 & 3.10 & 10.91 & 0.43 & 6.30 & 12.26 & 1.69 & 3.55 & 13.61 &  $-$ 0.06 &  $-$ 0.21 & 14.96 &  $-$ 4.81 &  $-$ 7.26 & 16.31 &  $-$ 5.33 &  $-$ 10.90 \\ 
 9.59 &  $-$ 2.33 & 3.16 & 10.94 & 0.60 & 6.36 & 12.29 & 1.68 & 3.39 & 13.64 &  $-$ 0.07 &  $-$ 0.32 & 14.99 &  $-$ 4.90 &  $-$ 7.28 & 16.34 &  $-$ 5.28 &  $-$ 11.09 \\ 
 9.62 &  $-$ 2.26 & 3.24 & 10.97 & 0.74 & 6.38 & 12.32 & 1.66 & 3.29 & 13.67 &  $-$ 0.09 &  $-$ 0.41 & 15.02 &  $-$ 4.96 &  $-$ 7.34 & 16.37 &  $-$ 5.24 &  $-$ 11.28 \\ 
 9.65 &  $-$ 2.18 & 3.32 & 11.00 & 0.87 & 6.42 & 12.35 & 1.64 & 3.23 & 13.70 &  $-$ 0.11 &  $-$ 0.47 & 15.05 &  $-$ 5.01 &  $-$ 7.46 & 16.40 &  $-$ 5.17 &  $-$ 11.53 \\ 
 9.68 &  $-$ 2.12 & 3.38 & 11.03 & 1.00 & 6.79 & 12.38 & 1.62 & 3.23 & 13.73 &  $-$ 0.14 &  $-$ 0.54 & 15.08 &  $-$ 5.07 &  $-$ 7.64 & 16.43 &  $-$ 5.12 &  $-$ 11.66 \\ 
 9.71 &  $-$ 2.04 & 3.46 & 11.06 & 1.15 & 6.66 & 12.41 & 1.59 & 3.20 & 13.76 &  $-$ 0.18 &  $-$ 0.66 & 15.11 &  $-$ 5.12 &  $-$ 7.83 & 16.46 &  $-$ 5.07 &  $-$ 11.71 \\ 
 9.74 &  $-$ 1.98 & 3.53 & 11.09 & 1.25 & 6.50 & 12.44 & 1.56 & 3.03 & 13.79 &  $-$ 0.21 &  $-$ 0.78 & 15.14 &  $-$ 5.17 &  $-$ 7.97 & 16.49 &  $-$ 5.00 &  $-$ 11.75 \\ 
 9.77 &  $-$ 1.92 & 3.60 & 11.12 & 1.34 & 6.32 & 12.47 & 1.53 & 2.86 & 13.82 &  $-$ 0.26 &  $-$ 0.94 & 15.17 &  $-$ 5.22 &  $-$ 8.13 & 16.52 &  $-$ 4.94 &  $-$ 11.77 \\ 
 9.80 &  $-$ 1.84 & 3.68 & 11.15 & 1.41 & 6.20 & 12.50 & 1.50 & 2.71 & 13.85 &  $-$ 0.32 &  $-$ 1.09 & 15.20 &  $-$ 5.26 &  $-$ 8.21 & 16.55 &  $-$ 4.89 &  $-$ 11.80 \\ 
 9.83 &  $-$ 1.77 & 3.75 & 11.18 & 1.46 & 6.15 & 12.53 & 1.47 & 2.55 & 13.88 &  $-$ 0.38 &  $-$ 1.20 & 15.23 &  $-$ 5.30 &  $-$ 8.28 & 16.58 &  $-$ 4.82 &  $-$ 11.85 \\ 
 9.86 &  $-$ 1.69 & 3.83 & 11.21 & 1.50 & 6.21 & 12.56 & 1.45 & 2.46 & 13.91 &  $-$ 0.46 &  $-$ 1.31 & 15.26 &  $-$ 5.35 &  $-$ 8.41 & 16.61 &  $-$ 4.77 &  $-$ 11.87 \\ 
 9.89 &  $-$ 1.60 & 3.90 & 11.24 & 1.51 & 6.27 & 12.59 & 1.44 & 2.41 & 13.94 &  $-$ 0.53 &  $-$ 1.38 & 15.29 &  $-$ 5.39 &  $-$ 8.55 & 16.64 &  $-$ 4.69 &  $-$ 11.85 \\ 
 9.92 &  $-$ 1.52 & 3.93 & 11.27 & 1.52 & 6.33 & 12.62 & 1.42 & 2.33 & 13.97 &  $-$ 0.60 &  $-$ 1.43 & 15.32 &  $-$ 5.43 &  $-$ 8.73 & 16.67 &  $-$ 4.63 &  $-$ 11.85 \\ 
 9.95 &  $-$ 1.43 & 4.00 & 11.30 & 1.52 & 6.33 & 12.65 & 1.40 & 2.25 & 14.00 &  $-$ 0.72 &  $-$ 1.48 & 15.35 &  $-$ 5.48 &  $-$ 8.92 & 16.70 &  $-$ 4.57 &  $-$ 11.92 \\ 
 9.98 &  $-$ 1.38 & 4.06 & 11.33 & 1.52 & 6.20 & 12.68 & 1.39 & 2.15 & 14.03 &  $-$ 0.81 &  $-$ 1.52 & 15.38 &  $-$ 5.50 &  $-$ 9.04 & 16.73 &  $-$ 4.50 &  $-$ 12.05 \\ 
 10.01 &  $-$ 1.35 & 4.13 & 11.36 & 1.50 & 5.97 & 12.71 & 1.38 & 2.06 & 14.06 &  $-$ 0.90 &  $-$ 1.56 & 15.41 &  $-$ 5.54 &  $-$ 9.10 & 16.76 &  $-$ 4.44 &  $-$ 12.24 \\ 
 10.04 &  $-$ 1.31 & 4.23 & 11.39 & 1.48 & 5.76 & 12.74 & 1.36 & 1.96 & 14.09 &  $-$ 1.02 &  $-$ 1.68 & 15.44 &  $-$ 5.56 &  $-$ 9.17 & 16.79 &  $-$ 4.38 &  $-$ 12.51 \\ 
 10.07 &  $-$ 1.26 & 4.28 & 11.42 & 1.47 & 5.61 & 12.77 & 1.33 & 1.89 & 14.12 &  $-$ 1.14 &  $-$ 1.86 & 15.47 &  $-$ 5.57 &  $-$ 9.30 & 16.82 &  $-$ 4.32 &  $-$ 12.79 \\ 
 10.10 &  $-$ 1.21 & 4.29 & 11.45 & 1.43 & 5.49 & 12.80 & 1.30 & 1.86 & 14.15 &  $-$ 1.24 &  $-$ 2.03 & 15.50 &  $-$ 5.59 &  $-$ 9.45 & 16.85 &  $-$ 4.26 &  $-$ 12.97 \\ 
 10.13 &  $-$ 1.15 & 4.25 & 11.48 & 1.41 & 5.50 & 12.83 & 1.25 & 1.81 & 14.18 &  $-$ 1.39 &  $-$ 2.22 & 15.53 &  $-$ 5.60 &  $-$ 9.55 & 16.88 &  $-$ 4.18 &  $-$ 13.11 \\ 
 10.16 &  $-$ 1.09 & 4.18 & 11.51 & 1.39 & 5.61 & 12.86 & 1.20 & 1.75 & 14.21 &  $-$ 1.51 &  $-$ 2.39 & 15.56 &  $-$ 5.61 &  $-$ 9.70 & 16.91 &  $-$ 4.12 &  $-$ 13.24 \\ 
 10.19 &  $-$ 1.04 & 4.15 & 11.54 & 1.38 & 5.65 & 12.89 & 1.14 & 1.69 & 14.24 &  $-$ 1.63 &  $-$ 2.54 & 15.59 &  $-$ 5.62 &  $-$ 9.79 & 16.94 &  $-$ 4.05 &  $-$ 13.40 \\ 
 10.22 &  $-$ 0.99 & 4.16 & 11.57 & 1.37 & 5.57 & 12.92 & 1.05 & 1.54 & 14.27 &  $-$ 1.79 &  $-$ 2.67 & 15.62 &  $-$ 5.63 &  $-$ 9.73 & 16.97 &  $-$ 3.95 &  $-$ 13.60 \\ 
 10.25 &  $-$ 0.96 & 4.23 & 11.60 & 1.36 & 5.49 & 12.95 & 0.98 & 1.41 & 14.30 &  $-$ 1.91 &  $-$ 2.77 & 15.65 &  $-$ 5.63 &  $-$ 9.63 & 17.00 &  $-$ 3.88 &  $-$ 13.77 \\ 
 10.28 &  $-$ 0.93 & 4.27 & 11.63 & 1.34 & 5.40 & 12.98 & 0.91 & 1.30 & 14.33 &  $-$ 2.06 &  $-$ 2.95 & 15.68 &  $-$ 5.62 &  $-$ 9.61 & 17.03 &  $-$ 3.80 &  $-$ 13.89 \\ 
 10.31 &  $-$ 0.89 & 4.17 & 11.66 & 1.34 & 5.33 & 13.01 & 0.83 & 1.17 & 14.36 &  $-$ 2.22 &  $-$ 3.18 & 15.71 &  $-$ 5.61 &  $-$ 9.63 & 17.06 &  $-$ 3.71 &  $-$ 13.97 \\ 
 10.34 &  $-$ 0.87 & 4.08 & 11.69 & 1.34 & 5.25 & 13.04 & 0.72 & 1.03 & 14.39 &  $-$ 2.35 &  $-$ 3.39 & 15.74 &  $-$ 5.61 &  $-$ 9.72 & 17.09 &  $-$ 3.62 &  $-$ 13.99 \\ 
 10.37 &  $-$ 0.86 & 3.92 & 11.72 & 1.34 & 5.21 & 13.07 & 0.64 & 0.99 & 14.42 &  $-$ 2.52 &  $-$ 3.71 & 15.77 &  $-$ 5.60 &  $-$ 9.86 & 17.12 &  $-$ 3.53 &  $-$ 14.03 \\ 
 10.40 &  $-$ 0.84 & 3.69 & 11.75 & 1.34 & 5.12 & 13.10 & 0.55 & 1.01 & 14.45 &  $-$ 2.65 &  $-$ 3.94 & 15.80 &  $-$ 5.59 &  $-$ 10.02 & 17.15 &  $-$ 3.39 &  $-$ 14.18 \\ 
 10.43 &  $-$ 0.84 & 3.50 & 11.78 & 1.34 & 4.97 & 13.13 & 0.45 & 0.99 & 14.48 &  $-$ 2.78 &  $-$ 4.14 & 15.83 &  $-$ 5.57 &  $-$ 10.13 & 17.18 &  $-$ 3.27 &  $-$ 14.39 \\ 
 10.46 &  $-$ 0.83 & 3.39 & 11.81 & 1.35 & 4.83 & 13.16 & 0.36 & 0.94 & 14.51 &  $-$ 2.95 &  $-$ 4.40 & 15.86 &  $-$ 5.56 &  $-$ 10.19 & 17.21 &  $-$ 3.15 &  $-$ 14.53 \\ 
 10.49 &  $-$ 0.83 & 3.41 & 11.84 & 1.37 & 4.70 & 13.19 & 0.25 & 0.83 & 14.54 &  $-$ 3.08 &  $-$ 4.58 & 15.89 &  $-$ 5.55 &  $-$ 10.19 & 17.24 &  $-$ 2.99 &  $-$ 14.73 \\ 
 10.52 &  $-$ 0.81 & 3.48 & 11.87 & 1.39 & 4.62 & 13.22 & 0.18 & 0.73 & 14.57 &  $-$ 3.23 &  $-$ 4.74 & 15.92 &  $-$ 5.55 &  $-$ 10.15 & 17.27 &  $-$ 2.86 &  $-$ 15.01 \\ 
 10.55 &  $-$ 0.78 & 3.63 & 11.90 & 1.42 & 4.54 & 13.25 & 0.12 & 0.63 & 14.60 &  $-$ 3.39 &  $-$ 4.91 & 15.95 &  $-$ 5.55 &  $-$ 10.15 & 17.30 &  $-$ 2.72 &  $-$ 15.37 \\ 
 10.58 &  $-$ 0.76 & 3.73 & 11.93 & 1.47 & 4.39 & 13.28 & 0.06 & 0.52 & 14.63 &  $-$ 3.52 &  $-$ 5.09 & 15.98 &  $-$ 5.55 &  $-$ 10.21 & 17.33 &  $-$ 2.57 &  $-$ 15.84 \\ 
 10.61 &  $-$ 0.71 & 3.80 & 11.96 & 1.50 & 4.23 & 13.31 & 0.02 & 0.44 & 14.66 &  $-$ 3.69 &  $-$ 5.36 & 16.01 &  $-$ 5.55 &  $-$ 10.25 & 17.36 &  $-$ 2.45 &  $-$ 16.56 \\ 
 10.64 &  $-$ 0.64 & 3.85 & 11.99 & 1.53 & 4.08 & 13.34 &  $-$ 0.01 & 0.34 & 14.69 &  $-$ 3.82 &  $-$ 5.57 & 16.04 &  $-$ 5.54 &  $-$ 10.31 & 17.39 &  $-$ 2.28 &  $-$ 17.13 \\ 
 10.67 &  $-$ 0.56 & 3.94 & 12.02 & 1.56 & 3.95 & 13.37 &  $-$ 0.03 & 0.22 & 14.72 &  $-$ 3.94 &  $-$ 5.78 & 16.07 &  $-$ 5.54 &  $-$ 10.44 & 17.42 &  $-$ 2.15 &  $-$ 16.85 \\ 
 10.70 &  $-$ 0.43 & 4.17 & 12.05 & 1.59 & 3.87 & 13.40 &  $-$ 0.03 & 0.11 & 14.75 &  $-$ 4.10 &  $-$ 6.04 & 16.10 &  $-$ 5.52 &  $-$ 10.55 & 17.45 &  $-$ 2.03 &  $-$ 16.86 \\ 
 10.73 &  $-$ 0.33 & 4.41 & 12.08 & 1.61 & 3.86 & 13.43 &  $-$ 0.04 & 0.03 & 14.78 &  $-$ 4.21 &  $-$ 6.21 & 16.13 &  $-$ 5.50 &  $-$ 10.59 & 17.48 &  $-$ 1.86 &  $-$ 16.85 \\ 
 10.76 &  $-$ 0.21 & 4.74 & 12.11 & 1.62 & 3.90 & 13.46 &  $-$ 0.04 & 0.01 & 14.81 &  $-$ 4.33 &  $-$ 6.45 & 16.16 &  $-$ 5.48 &  $-$ 10.64 & 17.51 &  $-$ 1.73 &  $-$ 16.83 \\ 
 10.79 &  $-$ 0.06 & 5.32 & 12.14 & 1.64 & 3.96 & 13.49 &  $-$ 0.04 & 0.01 & 14.84 &  $-$ 4.45 &  $-$ 6.71 & 16.19 &  $-$ 5.47 &  $-$ 10.68  \\ 
 10.82 & 0.05 & 5.71 & 12.17 & 1.66 & 3.97 & 13.52 &  $-$ 0.04 & 0.01 & 14.87 &  $-$ 4.54 &  $-$ 6.89 & 16.22 &  $-$ 5.45 &  $-$ 10.68 \\
 
    \end{tabular}
\end{table*}}

\section{Summary} \label{sec:sum}

Following Paper\,{\rm I}, by combining $\sim$ 0.7 million high-quality common stars in LAMOST DR7 with the SCR method, we obtain $G-G_{\rm RP}$ and $G_{\rm BP}-G_{\rm RP}$ color corrections as a function of $G$ magnitude for $9.5 < G < 17.5$ to sub-mmag precision for \textit{Gaia} EDR3. 
Our results confirm the improvements in the calibration process of the EDR3. 
The color term of the $G_{\rm BP}-G_{\rm RP}$ for bright stars is removed.
The discontinuity caused by the changes of instrument configurations is significantly reduced. 
Yet modest systematic trends with $G$ magnitude are detected.

By comparing with the work of \citet{yl}, we find that our color corrections of individual colors are subject to systematic errors in reddening correction with the SFD map. In the cases of color-color diagram, our corrections still achieve an unprecedented sub-mmag precision. Our work could be beneficial to studies where a high-precision color-color diagram is required, including 
estimates of \textit{Gaia} photometric metallicities and discrimination between binaries and single stars.

\acknowledgments

We acknowledge the anonymous referee for his/her valuable comments that improve the quality of this paper significantly. This work is supported by National Key Research and Development Program of China (NKRDPC) under grant numbers 2019YFA0405503, 2019YFA0405504, and 2016YFA0400804, National Science Foundation of China (NSFC) under grant numbers 11603002, 11988101, and 113300034, and Beijing Normal University grant No. 310232102.

This work has made use of data products from the Guoshoujing Telescope (the Large Sky Area Multi-Object Fiber Spectroscopic Telescope, LAMOST). LAMOST is a National Major Scientific Project built by the Chinese Academy of Sciences. Funding for the project has been provided by the National Development and Reform Commission. LAMOST is operated and managed by the National Astronomical Observatories, Chinese Academy of Sciences. This work has made use of data from the European Space Agency (ESA) mission {\it Gaia} (\url{https://www.cosmos.esa.int/gaia}), processed by the {\it Gaia} Data Processing and Analysis Consortium (DPAC, \url{https://www.cosmos.esa.int/web/gaia/dpac/consortium}). Funding for the DPAC has been provided by national institutions, in particular the institutions participating in the {\it Gaia} Multilateral Agreement.


\clearpage

\bibliography{sample63}{}
\bibliographystyle{aasjournal}

\end{document}